\documentclass[twoside,fleqn]{article}
\usepackage{epsf}
\usepackage{espcrc2}
\newcommand{\disable}[1]{}
\newcommand{\journal}[5]{#1;\textsl{ #2} \textbf{#3 }(#4) #5}
\newcommand{\ba}{\begin{eqnarray}}
\newcommand{\ea}{\end{eqnarray}}
\newcommand{\be}{\begin{equation}}
\newcommand{\ee}{\end{equation}}
\newcommand{\phd}{\phi^{\dagger}}
\newcommand{\mdm}{M^{\dagger}M}

\renewcommand{\dh}{\Delta\mathcal{H}}

\newcommand{\dtt}{\frac{d\tau}{2}}
\newcommand{\dttn}{\frac{d\tau}{2n}}
\newcommand{\dtn}{\frac{d\tau}{n}}

\title{Implementing The Generalised Hybrid Monte-Carlo Algorithm}
\author{Z. Sroczynski
\address{Department of Physics and Astronomy, 
University of Edinburgh, Edinburgh EH9 3JZ, Scotland},
S. M. Pickles $^{\mathrm{\alph{address}}}$ and 
S. P. Booth 
\address{EPCC, University of Edinburgh, Edinburgh EH9 3JZ, Scotland}
(UKQCD Collaboration)\thanks{Presented by Z. Sroczynski}
}

\begin{document}

\begin{abstract}
UKQCD's dynamical fermion project 
uses the Generalised Hybrid Monte Carlo (GHMC) algorithm 
to generate QCD gauge configurations 
for a non-perturbatively $O(a)$ improved Wilson action
with two degenerate sea-quark flavours.
We describe our implementation of the algorithm on the Cray-T3E, 
concentrating on issues arising from 
code verification and performance optimisation, such as
parameter tuning, 
reversibility,
the effect of precision,
the choice of matrix inverter, 
and the behaviour of different molecular dynamics integration schemes.         
\end{abstract}

\maketitle

\section{Introduction}
First
physics results from the UKQCD dynamical fermions project are presented 
elsewhere \cite{mauro}; here we
discuss issues arising from the implementation of GHMC \cite{ghmc} in 
Fortran90 on the Cray-T3E, such as the
optimisation of the code to take full advantage of the Cray-T3E
architecture, and interesting aspects of the algorithm's behaviour
demonstrated in the course of the code verification process.

\subsection{Notation}
Gauge fields are denoted $U$ and their canonically congugate
momentum fields $P$. The HMC hamiltonian is
\be
\mathcal{H} = T(P) + S_g  + S_f
\ee
where $S_g$ is the plaquette pure gauge action
and the fermionic part of the action is
\be
S_f=\phd(\mdm)^{-1}\phi -2\sum_{even\;x}\ln\det A_x
\ee
$S_f$ involves
pseudofermion fields $\phi$ and the odd-even preconditioned fermion matrix
\be
M_{xy} = A_{xx} - \kappa^2D_{xz}A_{zz}^{-1}D_{zy}
\ee
where $D$ is the usual Wilson hopping matrix 
and $A$ is the (Sheikoloslami-Wohlert) clover term.

\section{Code Improvements}

Since our original Fortran implementation, hardware specific code optimisation
and algorithmic improvements
have seen the CPU time required to complete one trajectory reduced
by a factor of  10 to 20, depending on lattice size, $\beta$
and $\kappa_{sea}$.

\subsection{Cray-T3E Architectural Factors}
Performance on the Cray-T3E is dominated by memory bandwidth. 
\disable{
Although the memory system fo the T3E is much improved relative to the
T3D (streams, S-cache, E-registers) the faster processors mean that
memory optimisations are more important.
}
The fact that the processors support multiple 
instruction issue coupled with the increased complexity of the memory
system makes it difficult to take full advantage
of the machine's capabilities with even highly optimised
Fortran code, so key routines {\em must} be written in assembler.
As a by-product of the tender process, we obtained
a set of highly optimised fermion matrix   multiplication routines.
We have re-written a number
of additional routines in assembler and now less
than 25\% of the run-time is spent executing Fortran code.

Using   32-bit instead 
of 64-bit floating point numbers to represent the fields
improves speed by a factor of 1.7.
This must be weighed against degradation of the acceptance rate,
the reversibility and the accuracy
of calculations of $\mathcal{H}$ 
(the correctness of the algorithm depends on the accurate computation of
the difference of two extensive quantities which are constrained to have
nearly identical magnitudes -- eventually, \textsl{i.e.} as the lattice volume
approaches the reciprocal of the 32-bit machine epsilon, the problem
becomes intractable). We have taken special care here, computing
energy differences site by site and performing summations in higher
precision. 

\subsection{Algorithmic Improvements}

Using BiCGStab \cite{bicgstab} 
instead of Conjugate Gradient to perform the inversions yielded
a $\kappa$ dependent saving of about 40\%. 

A simple observation which does not seem to be widely known
is that the vectors $r$ and $s$ in BiCGStab (in the notation of
\cite{bicgstab}) can and should be overlapped in memory, saving
one vector's worth of storage and two memory copies per iteration.

A further saving of one solve per trajectory
can be made by observing that at the start of a trajectory
the solution to $M^\dagger Y = \phi$ is known exactly 
(we initialise the $\phi$ fields by
the heatbath $\phi = M^\dagger \eta$ for Gaussian noise $\eta$).

There are special considerations when using the Clover action.
A similarity transformation reduces the $12\times  12$
matrix to two $6\times 6$ matrices \cite{stephenson}.  This
halves the memory requirement and doubles the speed
of our inverse multiply routine. Instead of storing $A^{-1}$
explicitly, we store the $L^\dagger D L$ decomposition of $A_{ee}$; 
this saves memory, is just as efficient, more robust and 
makes the computation of $\det A$
(in $S_f$) trivial.

Left-preconditioning the BiCGStab
by $A_{ee}^{-1}$ reduces the number of iterations required for 
convergence by some 15\%. Using $A_{ee}^{-1}$ as a central preconditioner
with CG saves about 25\% in terms of iterations. 

We do the exponentiation of the conjugate momenta
(as required in (\ref{int_t_p})) \textit{exactly} (up to machine precision).  
Our first implementation accomplished this by diagonalising the
3 by 3 matrices using library routines. 
We improved speed by a factor $>6$ 
with a method \cite{pieceofpaper}
which exploits the Cayley-Hamilton theorem and requires the solution to
a Vandermonde system.

\section{Integration Schemes}
\label{ints} 
The numerical integration of $U$ and $P$
forward in ``molecular dynamics time'' $\tau$ 
can be represented \cite{sex} as an evolution
operator $T(d\tau)$ 
\be
T(d\tau):\left( \begin{array}{c}U(\tau)\\P(\tau) \end{array} \right)\rightarrow
\left( \begin{array}{c}U(\tau+d\tau)\\P(\tau+d\tau) \end{array}\right)
\ee
In general $T$ is composed of two operators; $T_U$ and 
$T_P$ where
\ba
\label{int_t_p}
T_P(d\tau):U &\rightarrow & e^{id\tau P}U\\
\label{int_t_u}
T_U(d\tau):P&\rightarrow & P-id\tau\,\partial_U (S_g+S_f)
\ea

Being a numerical integrator, $T$ does not conserve $\mathcal{H}$ exactly
but introduces an error $\dh$. It is expected theoretically
\cite{si}\cite{scale}
that $\dh$ grows as a power of the timestep
$d\tau$. Verifying that this occurs with our code provides a strong test that 
we have implemented the equations of motion correctly.

We introduce three integration schemes;
\ba
T_1(d\tau) &=& T_P(d\tau)T_U(d\tau)\\
T_2(d\tau) &=& T_P(\dtt)T_U(d\tau)T_P(\dtt)\\
T_3(d\tau) &=& T_P(\frac{a}{2}d\tau) 
                    T_U(ad\tau)T_P(\frac{a+b}{2}d\tau)T_U(bd\tau)\nonumber\\
&&     T_P(\frac{a+b}{2}d\tau)T_U(ad\tau)T_P(\frac{a}{2}d\tau)
\ea
where $a = 1/(2 - 2^{1/3})$, and $b = -2^{1/3} / (2 - 2^{1/3})$.
The expectation \disable{\cite{moresex}} is that $T_1$,
$T_2$ and $T_3$ cause 
$\dh$ to vary as $d\tau^2$, $d\tau^3$ and $d\tau^5$. We
find that for some range of $d\tau$ where rounding errors do not dominate,
the slopes of plots of $\log \dh$ vs. $\log d\tau$ are respectively 
$1.982\pm 0.004$, 
$3.053\pm 0.002$ and $5.056 \pm 0.006$. Similar results were obtained
using just the pure gauge action and
the unimproved Wilson action. We can conclude that the equations of
motion have been correctly implemented.

The computationally costly part of the the evolution is the evaluation of
$\partial_US_f$ (since it involves the inversion of $\mdm$) in
(\ref{int_t_u}), so we split this into 
\ba
T_{pg}(d\tau):P&\rightarrow & P-id\tau\,\partial_US_g\\
T_f(d\tau):P&\rightarrow & P-id\tau\,\partial_U S_f
\ea
and form the integrator
\ba
\label{integrator}
T(d\tau)&=&T_f(\dtt)\left[T_{pg}(\dttn)T_P(\dtn)\right.\nonumber\\
&&\left.T_{pg}(\dttn)\right]^nT_f(\dtt)
\ea

Increasing $n$ means that $\dh$ is reduced without the additional
expense of evaluating $\partial_US_f$ more often.
We find that increasing $n$
above 2 does not have a great effect on the acceptance rate, and indeed
as the integration demands more computational operations per timestep,
the accumulation of rounding errors has a deleterious effect on reversibilty
and eventually $\dh$. Therefore we  use (\ref{integrator}) with
$n\le 2$ in production. 
\section{Estimation of $\kappa_{crit}$}

Motivated by the idea that a dynamical gauge configuration behaves with some
scaling behaviour near a critical point 
one forms the ansatz

\be
N_{CG} \propto \left(\frac{1}{\kappa}-\frac{1}{\kappa_{crit}}\right)^{\delta}
\ee
where $N_{CG}$ is the number of CG iterations required to invert $\mdm$
to some given accuracy. Then one expects that a plot
of $\log(N_{CG})$ against $\log(1/\kappa - 1/\kappa_{crit})$ would be
linear. By using various guesses for $\kappa_{crit}$ one can 
estimate its true value as the one that yields a straight line. This is shown
in figure \ref{iters} for the configurations at $\beta = 5.2$, $\kappa =
0.136$, $c_{SW}=1.72$ on a $12^3\times 24$ lattice.
{
\vspace{-0.275in}
\setlength{\unitlength}{1in}
\begin{center}
\begin{figure}[ht]
\begin{picture}(3.0,2.2)
\put(-0.5,-0.2){\begin{picture}(2.5,2.5)\put(-0.1,-0.7){\epsfxsize=3.5in 
\epsfbox[10 30 560 590]{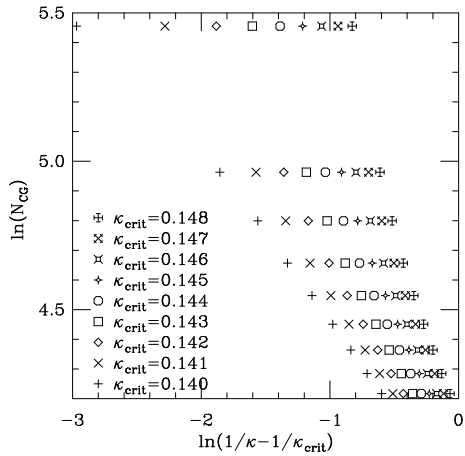}}\end{picture}}
\end{picture}
\caption{\label{iters}Estimation of $\kappa_{crit}$.}
\end{figure}
\end{center}
\vspace{-0.275in}
}
The straightest line lies between $\kappa_{crit}=0.140$ and
$\kappa_{crit}=0.141$. so we can perform a detailed linear fit in this range
(the result from subsequent spectroscopy is 0.14033(3)
\cite{mauro}). This is a useful technique since many 
solves are performed in a HMC run (strictly one should only count the
first solve of a trajectory, since this is the only one that is
guaranteed to be done on a physical configuration) so a statistically
significant value for $N_{CG}$ is easily obtained.
One then has a quick and reasonable accurate preliminary estimate of
$\kappa_{crit}$.

\section{Autocorrelation Analysis}

For the $12^3\times 24$ lattice at $\beta = 5.2$ we estimate the following
for the  exponential and integrated plaquette autocorrelation times
 $\tau_{exp}$ and $\tau_{int}$:\\
\begin{center}
\begin{tabular}{lccccc}
\hline
$\kappa$&0.136&0.137&0.138&0.139&0.1395\\ \hline
$\tau_{exp}$&17&28&40&36&50\\
$\tau_{int}$&20&34&36&45&64\\ \hline
\end{tabular}
\end{center}

\section*{Acknowledgements}
We gratefully acknowledge the support of PPARC in providing funds for the
Cray-T3E under grant GR/L22744.

Big shout going out to:
J. C. Sexton, B. J. Pendleton
and D. S. Henty for helpful advice and discussions,
and B\'{a}lint Jo\'{o} for writing some code for us.

\end{document}